\documentclass[10pt,journal,letterpaper]{IEEEtran}
\usepackage[T1]{fontenc}
\usepackage{newtxtext}
\usepackage{amsmath}
\usepackage{newtxmath}
\usepackage{booktabs,array,balance,cite,url}
\usepackage{tikz,pgfplots}
\pgfplotsset{compat=1.18}
\definecolor{paperblack}{HTML}{333333}
\definecolor{paperblue}{HTML}{0072B2}
\definecolor{paperred}{HTML}{D55E00}
\definecolor{papergreen}{HTML}{009E73}
\newcommand{\vect}[1]{\boldsymbol{#1}}
\newcommand{\T}{\mathsf{T}}
\newcommand{\diag}{\operatorname{diag}}
\newcommand{\rank}{\operatorname{rank}}

\newcommand{\SAR}{\mathrm{SAR}}
\begin{document}
\title{When Weak Reports Matter: Staged Anchored Fusion for Cooperative UAV Sensing}
\author{Xingqin~Lin\thanks{Xingqin Lin is with NVIDIA, USA (e-mail: xingqinl@nvidia.com).}}
\maketitle
\begin{abstract}
Local multipath rejection can erase evidence needed for cooperative sensing. We propose staged anchored recovery: preserve strong-only confirmations, then query compatible weak reports using unused strong anchors. For any number of sensing nodes, we prove lossless residual screening and derive corroboration and bidirectional cost laws. In 1,024 five-UAV drops, recovery adds 30 matched targets and three false outputs over strict consensus, matching one-pass anchored confirmation's detection counts while reducing weak uploads by 97.4\%. Equal-sized cue/report records yield 4.5\% less payload than uploading all eligible reports. Independent validation recovers four additional targets with no observed false outputs.
\end{abstract}

\section{Introduction}
\IEEEPARstart{A}{target} rejected by one sensing receiver may be recoverable by the network if its evidence survives local processing. In integrated sensing and communication (ISAC) for unmanned aerial vehicle (UAV) monitoring, multipath rejection suppresses unreliable echoes, yet propagation and viewing angle can make one UAV strong at one transmit/receive point (TRP) and weak at another. Requiring multiple locally accepted reports can miss the target; forwarding every weak report increases association search and traffic. Recovery therefore depends on deciding which rejected reports deserve a second chance.

Wei \emph{et al.}~\cite{wei} establish positioning reference signal (PRS) sensing. Shi and Liu~\cite{shi} jointly identify line-of-sight (LOS) paths and associate targets under multipath; Tang \emph{et al.}~\cite{tang} combine false-removing association with UAV position/velocity estimation. Zhang \emph{et al.}~\cite{weighted} study multilevel weighted fusion for aerial localization. Resource design includes cooperative sensing-rate optimization across bands~\cite{perovic} and joint transmission/echo compression under fronthaul constraints~\cite{zhu}. Censored sensor decisions establish selective reporting~\cite{jiang}. We address selective recovery of locally rejected reports while preserving confirmed estimates and accounting for query and response traffic.

We propose \emph{staged anchored recovery} (SAR): lock objects confirmed by strong reports, then query compatible weak reports using unused strong anchors. For any number of TRPs and generic three-dimensional motion, we prove lossless screening, derive corroboration laws, and bound weak-eligibility sensitivity and bidirectional cost. Paired PRS-based UAV experiments share gates and fitting, isolating staged admission and reporting.

\section{Report Model and Staged Recovery}
\subsection{Strong and weak evidence}
Consider $K\geq2$ monostatic TRPs. A report $a$ from TRP $i(a)\in\{1,\ldots,K\}$ has a stable identifier and
\begin{equation}
\vect z_a=(r_a,\alpha_a,\epsilon_a,\nu_a)^\T,
\label{eq:report}
\end{equation}
where $r_a$, $\alpha_a$, $\epsilon_a$, and $\nu_a$ are range, global azimuth, elevation, and radial velocity (positive receding). It also carries TRP pose, reference epoch, quality, local-acceptance flag, and measurement-weight matrix $\Sigma_a$. Reports passing all local checks form the strong set $S_i$. A larger eligible set $E_i\supseteq S_i$ meets fixed lower quality thresholds; $W_i=E_i\setminus S_i$ is the weak set. Weak means \emph{locally rejected but still eligible}.

Let $\vect b_i$ be TRP $i$'s surveyed position and $\vect u_a=(\cos\epsilon_a\cos\alpha_a,\cos\epsilon_a\sin\alpha_a,\sin\epsilon_a)^\T$. Report $a$ implies position $\vect p_a=\vect b_{i(a)}+r_a\vect u_a$, with range/angle Jacobian
\begin{equation}
J_a=\left[\vect u_a,\ r_a\frac{\partial\vect u_a}{\partial\alpha_a},\ r_a\frac{\partial\vect u_a}{\partial\epsilon_a}\right],
\label{eq:jacobian}
\end{equation}
with angles in radians. Propagating the report weights gives $C_a=J_a\Sigma_{a,1:3}J_a^\T+\varepsilon_C I$. A fixed geometric gate $G(a,b)$ between reports from distinct TRPs tests
\begin{equation}
(\vect p_a-\vect p_b)^\T(C_a+C_b+C_{\rm motion})^{-1}
(\vect p_a-\vect p_b)\leq\chi^2_3(0.9973).
\label{eq:gate}
\end{equation}
Here $\Sigma_{a,1:3}$ is the range/azimuth/elevation $3\times3$ block of $\Sigma_a$, $\chi^2_d(u)$ is the $u$-quantile with $d$ degrees of freedom, $\varepsilon_C$ regularizes position weights, and $C_{\rm motion}$ accounts for motion between epochs. The dimension three in~\eqref{eq:gate} concerns position, independently of $K$.

Diagonal report weights use
\begin{equation}
\sigma_{\xi,a}=\max\{\delta_{\xi,a}/\sqrt{12},a_{\xi,a}/\sqrt{\gamma_{{\rm eff},a}}\},
\quad\xi\in\{r,\alpha,\epsilon,\nu\}.
\label{eq:weights}
\end{equation}
The resolution floors $\delta_{\xi,a}$ and scales $a_{\xi,a}$ follow the receiver grid, PRS bandwidth, beamwidth, and slow-time aperture. Effective SNR $\gamma_{{\rm eff},a}$ combines observed SNR and local consistency scores; $\Sigma_a=\diag(\sigma_{r,a}^2,\sigma_{\alpha,a}^2,\sigma_{\epsilon,a}^2,\sigma_{\nu,a}^2)$. Section~IV specifies the experimental values.

Each candidate group contains $2\leq M\leq K$ reports, at most one per TRP, and passes every pair gate. \emph{Strict consensus} (SC) uses strong-only groups; \emph{permissive confirmation} (PC) uses eligible reports; \emph{anchored confirmation} (AC) additionally requires a strong member. AC fits and selects groups in one pass, allowing weak reports to augment or replace SC groups. SAR first locks SC outputs. All policies share fitting and exclusive association.

\subsection{Recovery after strict consensus}
\begin{figure}[t]
\centering
\begin{tikzpicture}[font=\footnotesize,>=stealth,
strong/.style={circle,draw=paperblue,fill=paperblue,minimum size=5pt,inner sep=0pt},
weak/.style={circle,draw=paperred,fill=white,minimum size=5pt,inner sep=0pt}]
\node[align=center] at (.8,.72) {1. Strict consensus};
\node[align=center] at (3.65,.72) {2. Residual recovery};
\node[align=center] at (6.5,.72) {3. All-weak group};
\node[strong] (a1) at (0,0) {};
\node[strong] (a2) at (1.6,0) {};
\draw[paperblue,dashed] (a1) -- (a2);
\node[below=3pt] at (a1) {TRP 1};
\node[below=3pt] at (a2) {TRP 2};
\node[draw=paperblue,rounded corners=2pt,text width=2.0cm,align=center,inner sep=3pt]
at (.8,-1.0) {Lock estimate\\and member IDs};
\node[strong] (b1) at (2.85,0) {};
\node[weak] (b2) at (4.45,0) {};
\draw[->,paperblack] (b1.north) to[bend left=25] node[above,font=\scriptsize] {cue / return} (b2.north);
\draw[paperblue,dashed] (b1) -- (b2);
\node[below=3pt] at (b1) {unused};
\node[below=3pt] at (b2) {weak};
\node[draw=papergreen,rounded corners=2pt,text width=2.0cm,align=center,inner sep=3pt]
at (3.65,-1.0) {Fit, check,\\append recovery};
\node[weak] (c1) at (5.7,0) {};
\node[weak] (c2) at (7.3,0) {};
\draw[paperred,dashed] (c1) -- (c2);
\node[below=3pt] at (c1) {weak};
\node[below=3pt] at (c2) {weak};
\node[draw=paperred,rounded corners=2pt,text width=2.0cm,align=center,inner sep=3pt]
at (6.5,-1.0) {No strong anchor:\\cannot confirm};
\end{tikzpicture}
\caption{Staged recovery for any $K\geq2$. Filled/open circles denote strong/eligible weak reports; dashed lines denote compatible gates. Locked members are retained; unused strong reports solicit weak evidence. All-weak groups cannot confirm an object.}
\label{fig:mechanism}
\end{figure}
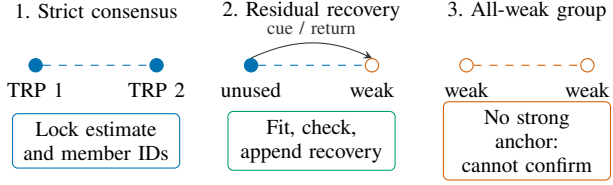

SAR performs the following operations (Fig.~\ref{fig:mechanism}).

\noindent\textbf{1) Confirm and lock.} Upload strong reports, run SC, and lock accepted estimates and member IDs. Locked reports cannot enter another group. Let $S^{\rm res}$ be the unused strong reports.

\noindent\textbf{2) Cue residual anchors.} Each $s\in S^{\rm res}$ queries the other $K-1$ TRPs for eligible weak reports passing $G(s,w)$. Deduplicate replies by stable ID, yielding
\begin{equation}
W^{\rm cue}=\{w\in W:\exists s\in S^{\rm res},\ G(s,w)=1\},
\label{eq:cue}
\end{equation}
where $W=\bigcup_{i=1}^K W_i$.

\noindent\textbf{3) Confirm residual groups.} Enumerate all-pair-compatible groups from $S^{\rm res}\cup W^{\rm cue}$ containing at least two TRPs and a residual anchor. Apply the shared fitter and exclusive selection, then append accepted estimates to the locked outputs.

For a group of $M$ reports, the shared fitter estimates $\vect x=(p_x,p_y,p_z,v_x,v_y,v_z)^\T$. Set $\vect v=(v_x,v_y,v_z)^\T$, $\vect d_a=\vect p+\vect v(t_a-t_0)-\vect b_{i(a)}$, and $R_a=\|\vect d_a\|$, where $t_a$ and $t_0$ are report and fusion epochs. The forward model is
\begin{equation}
\begin{aligned}
h_a(\vect x)=\big(&R_a,\operatorname{atan2}(d_{a,y},d_{a,x}),\\
&\operatorname{atan2}(d_{a,z},\sqrt{d_{a,x}^2+d_{a,y}^2}),\vect d_a^\T\vect v/R_a\big)^\T.
\end{aligned}
\label{eq:forward}
\end{equation}
With $\Sigma_a=L_aL_a^\T$, the fitter minimizes
\begin{equation}
\sum_{a\in\mathcal G}\sum_{k=1}^4\rho_{1.5}\!\left([L_a^{-1}\{\vect z_a-h_a(\vect x)\}]_k\right),
\label{eq:fit}
\end{equation}
where $\mathcal G$ is the group, angular residuals are wrapped, and $\rho_{1.5}$ is Huber loss. Let $D_{\mathcal G}=[D_p,D_v]$ stack $L_a^{-1}\partial h_a/\partial\vect x$ in position/velocity columns. The nominal local residual count $n_{\mathcal G}=4M-\rank D_{\mathcal G}$ becomes $4M-6$ for full rank. Gate maximum absolute whitened residual at four, squared whitened residual sum $Q$ at $\chi^2_{n_{\mathcal G}}(0.999)$, and fitted speed at a prescribed maximum $v_{\max}$. Rank groups by decreasing strong support, decreasing total support, increasing $Q/n_{\mathcal G}$, decreasing quality, then stable IDs; selection forbids member reuse. Profile position via $B_v=(I-D_pD_p^\dagger)D_v$, with $\dagger$ the pseudoinverse. Full velocity requires rank three and singular-value ratio at most 100 for $B_v$. Rank-deficient fits use a minimum-norm representative and report observable velocity projections with their basis. When $t_a=t_0$, two nonparallel TRPs give rank five and $n_{\mathcal G}=4M-5$; target confirmation remains independent of full velocity availability.

\section{What Anchoring Preserves and Costs}
\subsection{Lossless screening and retention}
\noindent\textbf{Proposition 1 (lossless residual screening).}
For fixed reports and pair gates, let $\mathcal H(F)$ contain every candidate group in $F$ with a residual strong member. Then
\begin{equation}
\mathcal H(S^{\rm res}\cup W^{\rm cue})=
\mathcal H(S^{\rm res}\cup W).
\label{eq:lossless}
\end{equation}
\emph{Proof:} The left family is a subset of the right. Choose a strong member $s$ in any right-hand group. Every weak member passes $G(s,w)$ and is returned by~\eqref{eq:cue}; the entire group belongs to the left family. \hfill$\square$

With unchanged measurements, weights, stable IDs, and settings, complete cueing preserves residual outputs relative to uploading all weak reports after identical SC locks, for any velocity dimension. One-pass AC has no prior SC locks. SAR preserves locked outputs; appending $r$ gives $0\leq\Delta M\leq r$ and $\Delta F=r-\Delta M\geq0$ under maximum-cardinality truth matching, where $\Delta M$ and $\Delta F$ are changes in matched and unmatched counts.

\subsection{Corroboration with an arbitrary number of views}
Consider one hypothesized object with nested strong/eligible events $S_i\subseteq E_i$, having probabilities $p_i\leq q_i$. Let $C_K$, $A_K$, and $R_K$ denote the SC, AC, and PC evidence probabilities. For independent views, anchored corroboration is
\begin{equation}
A_K=1-\prod_{i=1}^K(1-p_i)
-\sum_{i=1}^K p_i\prod_{j\ne i}(1-q_j).
\label{eq:hetero}
\end{equation}
The excluded events are no strong view, and exactly one strong view with every other view ineligible. For identical probabilities $p_i=p$ and $q_i=q$, direct counting gives
\begin{align}
C_K&=1-(1-p)^K-Kp(1-p)^{K-1},\nonumber\\
A_K&=1-(1-p)^K-Kp(1-q)^{K-1},\label{eq:law}\\
R_K&=1-(1-q)^K-Kq(1-q)^{K-1}.\nonumber
\end{align}
The anchored recovery gain over SC is exactly the event of one strong view corroborated by at least one weak view:
\begin{equation}
A_K-C_K=Kp\big[(1-p)^{K-1}-(1-q)^{K-1}\big].
\label{eq:gain}
\end{equation}
The probability removed from PC is the all-weak event
\begin{equation}
R_K-A_K=(1-p)^K-(1-q)^K-K(q-p)(1-q)^{K-1}.
\label{eq:excluded}
\end{equation}
With objectwise association, compatible reports, and successful fits, both SAR and AC realize $A_K$: multiple-strong objects lock first under SAR, and one-strong objects are recovered.

For false evidence, write $p=\alpha$ and $q=\beta$. A union bound over distinct TRP pairs gives
\begin{equation}
A_K^{\rm false}\leq\min\!\left\{1,{K\choose2}(2\alpha\beta-\alpha^2)\right\}.
\label{eq:false}
\end{equation}
For small probabilities and fixed $K$, anchored false corroboration scales with $\alpha\beta$, while permissive false corroboration scales with $\beta^2$. The selective advantage grows when strong false evidence is substantially rarer than eligible false evidence.

\noindent\textbf{Proposition 2 (weak-eligibility sensitivity bound).}
For independent identical views, hold $p$ fixed and let $0\leq p\leq q_0<q_1<1$, $q_0>0$. For every $K\geq2$,
\begin{equation}
\frac{p}{q_1}\Delta R_K\leq\Delta A_K\leq\frac{p}{q_0}\Delta R_K,
\label{eq:expansion}
\end{equation}
where $\Delta A_K=A_K(q_1)-A_K(q_0)$ and similarly for $\Delta R_K$.

\emph{Proof:} Differentiating~\eqref{eq:law} gives
\begin{equation}
\begin{aligned}
A_K'(q)&=K(K-1)p(1-q)^{K-2},\\
R_K'(q)&=K(K-1)q(1-q)^{K-2}.
\end{aligned}
\label{eq:derivatives}
\end{equation}
Thus $A_K'=(p/q)R_K'$. Integrating over $[q_0,q_1]$ and bounding $p/q$ by its endpoint values proves~\eqref{eq:expansion}. \hfill$\square$

At eligibility probability $q$, the marginal anchored corroboration gain is $p/q$ times the permissive gain. For false evidence with $\alpha=10^{-4}$ and $\beta$ expanded from 0.005 to 0.01, anchoring admits only 1--2\% of that increment for any $K$, with the strong threshold fixed.

\begin{figure}[t]
\centering
\begin{tikzpicture}
\begin{axis}[at={(0,0)},anchor=south west,width=.49\columnwidth,height=1.62in,
xmin=2,xmax=12,xtick={2,4,6,8,10,12},ymin=.4,ymax=1.02,grid=major,grid style={gray!15},axis lines=left,
xlabel={Number of TRPs $K$},ylabel={Target corroboration probability},
tick label style={font=\scriptsize},label style={font=\scriptsize},
legend style={font=\scriptsize,draw=none,fill=none,at={(.5,1.03)},anchor=south},legend image post style={xscale=.45},
legend columns=1,legend cell align=left,domain=2:12,samples=11]
\addplot[paperblack,thick,mark=* ,mark size=1pt] coordinates {(2,0.4225) (3,0.71825) (4,0.87351875) (5,0.9459775) (6,0.977678203125) (7,0.9909924984375) (8,0.99642916902344) (9,0.99860383725781) (10,0.9994601128751) (11,0.99979310894849) (12,0.99992131243674)};\addlegendentry{SC}
\addplot[paperblue,thick,mark=triangle*,mark size=1.2pt] coordinates {(2,0.6825) (3,0.91325) (4,0.97621875) (5,0.9931025) (6,0.997865578125) (7,0.9993047796875) (8,0.99976592777344) (9,0.99991968507031) (10,0.99997216464463) (11,0.99999030385376) (12,0.99999661403268)};\addlegendentry{AC}
\addplot[paperred,thick,dashed,mark=square*,mark size=1pt] coordinates {(2,0.7225) (3,0.93925) (4,0.98801875) (5,0.9977725) (6,0.999601328125) (7,0.9999305171875) (8,0.99998812527344) (9,0.99999800094531) (10,0.99999966746494) (11,0.99999994521821) (12,0.9999999910475)};\addlegendentry{PC}
\end{axis}
\begin{semilogyaxis}[at={(.51\columnwidth,0)},anchor=south west,width=.49\columnwidth,height=1.62in,
xmin=2,xmax=12,xtick={2,4,6,8,10,12},ymin=5e-7,ymax=.003,grid=major,grid style={gray!15},axis lines=left,
xlabel={Number of TRPs $K$},ylabel={False corroboration probability},
tick label style={font=\scriptsize},label style={font=\scriptsize},
legend style={font=\scriptsize,draw=none,fill=none,at={(.5,1.03)},anchor=south},legend image post style={xscale=.45},
legend columns=1,legend cell align=left,domain=2:12,samples=11]
\addplot[paperblue,thick,mark=triangle*,mark size=1.2pt] coordinates {(2,9.8999999992771e-07) (3,2.9625010000161e-06) (4,5.910053999899e-06) (5,9.8252596869433e-06) (6,1.470076812528e-05) (7,2.0529278447042e-05) (8,2.7303538545349e-05) (9,3.5016344769459e-05) (10,4.3660541621316e-05) (11,5.3229021454582e-05) (12,6.3714724174253e-05)};\addlegendentry{AC: independent}
\addplot[paperblue,thick,dashed,mark=o,mark size=1.2pt] coordinates {(2,1.0890999999935e-05) (3,1.2666250900015e-05) (4,1.5319048599909e-05) (5,1.8842733718249e-05) (6,2.3230691312752e-05) (7,2.8476350602338e-05) (8,3.4573184690814e-05) (9,4.1514710292513e-05) (10,4.9294487459185e-05) (11,5.7906119309124e-05) (12,6.7343251756827e-05)};\addlegendentry{AC: shared event}
\addplot[paperred,thick,densely dotted,mark=square*,mark size=1pt] coordinates {(2,2.4999999999956e-05) (3,7.4749999999983e-05) (4,0.00014900187500005) (5,0.00024750936250002) (6,0.00037002805007813) (7,0.00051631536304646) (8,0.0006861305521838) (9,0.00087923468154574) (10,0.0010953906163502) (11,0.0013343630109396) (12,0.0015959182968177)};\addlegendentry{PC: independent}
\end{semilogyaxis}
\end{tikzpicture}
\caption{Corroboration model versus $K$. Left: target evidence, $p=0.65$, $q=0.85$. Right: false evidence, $\alpha=10^{-4}$, $\beta=0.005$. Solid/dashed blue denote independent/shared evidence ($\eta=0.1$).}
\label{fig:law}
\end{figure}
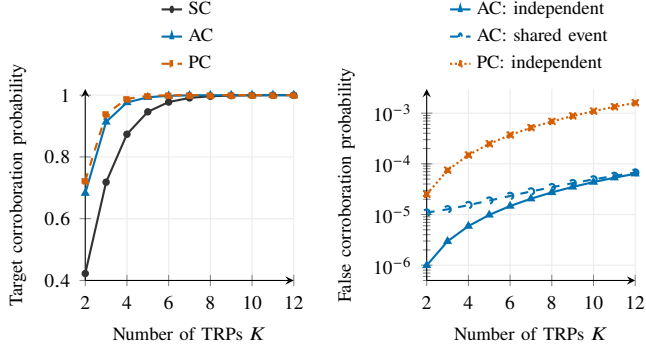

Dependence can be included by drawing one common uniform variable for all views with probability $\eta$ and independent variables otherwise. Thresholds $p$ and $q$ define the nested events. For any $K\geq2$,
\begin{equation}
A_{K,\eta}=\eta p+(1-\eta)A_K,
\label{eq:shared}
\end{equation}
with analogous expressions for $C_{K,\eta}$ and $R_{K,\eta}$. Here $\eta$ is the shared-event fraction. A common strong false event contributes $\eta\alpha$ regardless of $K$ (Fig.~\ref{fig:law}); adding TRPs cannot remove this shared component.

\subsection{Bidirectional reporting cost}
Let $N_s$ count uploaded strong reports, $N_w$ eligible weak reports, $N_a$ residual anchors, and $N_p$ deduplicated weak replies. Unicast querying uses $(K-1)N_a$ cues. With $b_r$ bits per report and $b_c$ per cue, AC with all eligible reports uploaded and SAR incur
\begin{align}
B_{\rm push}&=b_r(N_s+N_w),\nonumber\\
B_{\SAR}&=b_r(N_s+N_p)+b_c(K-1)N_a.\label{eq:cost}
\end{align}
For $N_a>0$, SAR reduces this bidirectional payload precisely when
\begin{equation}
\frac{b_c}{b_r}<\frac{N_w-N_p}{(K-1)N_a}.
\label{eq:break}
\end{equation}
This report/query model counts both directions before protocol framing and acknowledgments. When $N_a=0$, no query is sent and only $N_s$ reports are uploaded. Confirming first reduces the number of anchors, directly relaxing the cost boundary.

\section{PRS-Based UAV Evaluation}
\subsection{Configuration and experimental controls}
Following the UAV study context of 3GPP TR~38.765~\cite{study}, three monostatic TRPs at $(0,0,25)$, $(500,0,25)$, and $(250,433.01,25)$~m sense five UAVs per occupied drop; empty drops contain background only. UAVs are uniform in area within a central $120^\circ$ sector of radius $500/\sqrt3$~m, heights 25--300~m, horizontal speeds 0--50~m/s ($v_z=0$), and minimum separation 10~m. Target channels use aerial urban-macrocell propagation and reciprocal Option-2 subpath pairing. Small-UAV radar cross section has baseline $-12.81$~dBsm, lognormal spread 3.74~dB, mean-compensating offset, and three-standard-deviation upper cap. Each TRP has an independent three-reference-point urban-macrocell background; reciprocal sensing legs share the LOS draw. Geometry is fixed within the coherent processing interval (CPI), with evolving Doppler phase.

\begin{table}[t]
\caption{Common acquisition and receiver configuration}
\label{tab:setup}\centering\footnotesize
\begin{tabular}{@{}>{\raggedright\arraybackslash}p{.39\columnwidth}@{\hspace{.03\columnwidth}}>{\raggedright\arraybackslash}p{.58\columnwidth}@{}}
\toprule
Carrier / subcarrier spacing & 4 GHz / 30 kHz \\
TRP receive panel & $8\times8$ dual-slant positions; 128 ports \\
Element spacing & $0.5\lambda$ horizontal, $0.8\lambda$ vertical \\
PRS per occasion & Three one-port beamforming resources, four comb-2 symbols each \\
Slow time & 32 occasions, 4-ms spacing; nominal 128-ms CPI \\
Three-TRP partition & 92/92/88 RBs; 47.229 dBm per TRP \\
Aggregate transmit budget & 272 RBs and energy of one 52-dBm TRP \\
Noise / residual SI & 5-dB NF; white $-94$ dBm per 100 MHz \\
\bottomrule
\end{tabular}
\end{table}

Table~\ref{tab:setup} fixes acquisition for all policies. The aggregate PRS resource-element fraction is $(272/273)(12/112)(1/2)=5.3375\%$ of the carrier grid. Known-pilot per-RE channel acquisition, static-mean removal, Hann-windowed range/Doppler transforms, noise-model thresholding, and peak selection precede local checks. Strong reports require spatial/temporal explained-energy ratios at least 0.80/0.72, half-CPI power ratio at least 0.20, half-CPI direction/velocity differences at most $8^\circ$/1.25~m/s, delayed-companion rejection, and TX-signature consistency at least 0.85. Weak eligibility uses 0.50/0.45, 0.08, $16^\circ$/2.5~m/s, and 0.40, respectively. Delayed-companion rejection tests 6--100-m apparent-range separation, power ratio at most 0.025, and direction/velocity differences at most $8^\circ$/8~m/s.

In~\eqref{eq:weights}, $a_r=c/(2B)$, $a_{\alpha,\epsilon}=0.886/(7d_{h,v}\,2.355)$, and $a_\nu=\delta_\nu$; $B$ is PRS bandwidth and $d_{h,v}$ are wavelength-normalized element spacings. Grid floors are $\delta_r=c/(2N_r\Delta f)$, $\delta_{\alpha,\epsilon}=2^\circ$ in radians, and $\delta_\nu=\lambda/(2NT)$, with $N_r=4096$, $N=32$, and $T=4$~ms. Effective SNR is $\max\{\gamma_{\rm obs}\max(q_sq_t,0.05),1\}$, where $\gamma_{\rm obs}$ is noise-subtracted signal/noise energy ratio and $q_s,q_t$ are spatial/temporal scores. Reports share an epoch, so $C_{\rm motion}=0$; $\varepsilon_C=10^{-6}$~m$^2$. The horizontal-motion receiver estimates $(p_x,p_y,p_z,v_x,v_y)$ and retains nominal residual count $4M-5$. Its horizontal velocity test requires rank two and singular-value ratio at most 100 for $A_{H,aj}=d_{a,j}/(R_a\sigma_{\nu,a})$, $j\in\{x,y\}$. The speed gate uses $v_{\max}=60$~m/s; settings are fixed across policies.

We developed anchoring and staging using 1,024 occupied and 1,024 empty drops, termed the \emph{design population}. An independent \emph{validation population} contains 128 occupied and 128 empty drops. Parameters and implementation were fixed before examining validation results. All policies use the same underlying measurement reports within each drop, ensuring a paired comparison.

Scoring uses maximum-cardinality, minimum-distance one-to-one matching within 30~m. For $T_d$ truths, $D_d$ outputs, and $M_d$ matches, missed-target fraction is $(T_d-M_d)/T_d$ and type-2 false fraction (FA2) is $(D_d-M_d)/D_d$. FA2 is averaged over occupied drops with $D_d>0$. Intervals use 2,000 paired whole-drop bootstrap replicates; empty-drop alarm bounds are exact binomial bounds. Common-truth accuracy evaluates errors on identical matched truths.

\subsection{Detection, estimation, and reporting}
SAR recovers 30 truths over SC: misses fall from 194 to 164 and unmatched outputs rise from two to five (Table~\ref{tab:results}), recovering 15.5\% of SC's misses. AC and SAR both match 4,956 truths with five unmatched outputs; PC matches 4,956 but yields 77. Excluding all-weak groups removes 72 unmatched outputs here.

SAR preserves all 4,928 SC outputs, foregoing weak-report refinement. On 4,956 common matched truths, AC/SAR position-error P90 is 2.599/2.630~m; on 4,949 common truths with observable velocity, horizontal-velocity-error P90 is 0.290/0.301~m/s, respectively.

\begin{figure}[t]
\centering
\begin{tikzpicture}
\begin{axis}[at={(0,0)},anchor=south west,width=.49\columnwidth,height=1.67in,
ybar=2pt,bar width=6pt,ymin=0,ymax=230,axis lines=left,grid=major,grid style={gray!15},
xtick=data,xticklabels={SC,PC,AC,SAR},ylabel={Number of objects},
tick label style={font=\scriptsize},label style={font=\scriptsize},
legend style={font=\scriptsize,draw=none,fill=none,at={(.5,1.08)},anchor=south},legend image post style={xscale=.45},legend cell align=left,
nodes near coords,nodes near coords style={font=\tiny},enlarge x limits=.15]
\addplot[fill=paperblack!65,draw=paperblack] coordinates {(0,194) (1,164) (2,164) (3,164)};\addlegendentry{Missed truths}
\addplot[fill=paperred!75,draw=paperred] coordinates {(0,2) (1,77) (2,5) (3,5)};\addlegendentry{Unmatched outputs}
\end{axis}
\begin{axis}[at={(.51\columnwidth,0)},anchor=south west,width=.49\columnwidth,height=1.67in,
xmin=0,xmax=2.5,ymin=.88,ymax=1.25,axis lines=left,grid=major,grid style={gray!15},
xtick={0,1,2},ylabel={Normalized payload},xlabel={Cue/report bits $b_c/b_r$},
ytick={.9,1,1.1,1.2},scaled y ticks=false,
tick label style={font=\scriptsize},label style={font=\scriptsize},
legend style={font=\scriptsize,draw=none,fill=none,at={(.5,1.08)},anchor=south},legend image post style={xscale=.45},legend cell align=left,
domain=0:2.5,samples=51]
\addplot[paperblack,thick,densely dotted] {1};\addlegendentry{AC: push eligible}
\addplot[papergreen,thick] {13844/15176+x*644/15176};\addlegendentry{SAR}
\addplot[paperred,thick,dashed] {14176/15176+x*27618/15176};\addlegendentry{AC: cue all strong}
\end{axis}
\end{tikzpicture}
\caption{Design-population detection and reporting. Left: AC and SAR have equal detection counts. Right: payload relative to AC pushing all eligible reports, versus cue/report size. SAR cues 322 residual anchors; AC cue-all uses naive unicast and reaches 2.75 at unit ratio (off-scale).}
\label{fig:results}
\end{figure}
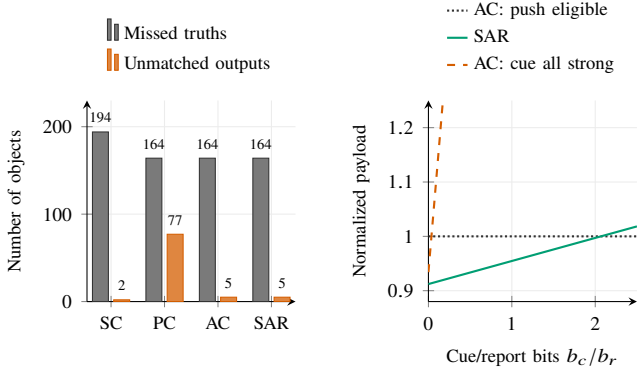

Reporting savings compare SAR with AC implemented by uploading all eligible reports. Both upload 13,809 strong records. AC sends 1,367 weak reports; SAR requests 35 (97.44\% fewer) via 644 cues from 322 residual anchors, after SC locks 13,487 strong members. The break-even ratio in~\eqref{eq:break} is 2.068. At $b_c=b_r$, 14,488 versus 15,176 record equivalents gives 4.53\% less modeled payload. AC can also cue: querying all strong reports returns 367 weak records with 27,618 unicast cues; its break-even ratio is 0.0362 (Fig.~\ref{fig:results}). These costs concern the specified reporting implementations.

The paired SAR--SC miss difference is $-0.586$ percentage points [${-0.801},{-0.371}$]; FA2 increases by 0.052 points [0,0.120]. Of 4,926 common matched truths, 4,925 retain the same locked output and one is reassociated by scoring. SAR velocity is available for 96.66\% of all truths. All policies have zero alarms in 1,024 design empty drops, giving a one-sided 95\% upper probability bound of 0.292\%.

\begin{table}[t]
\caption{Design population: 5,120 truths, 1,024 occupied drops. Brackets are pointwise 95\% intervals.}
\label{tab:results}\centering\footnotesize
\begin{tabular}{@{}lrrr@{}}
\toprule
Policy & Miss (\%) & FA2 (\%) & False outputs \\
\midrule
SC & 3.789 [3.301, 4.316] & 0.036 [0.000, 0.094] & 2 \\
PC & 3.203 [2.734, 3.691] & 1.285 [1.003, 1.578] & 77 \\
AC & 3.203 [2.734, 3.691] & 0.088 [0.016, 0.176] & 5 \\
SAR & 3.203 [2.734, 3.691] & 0.088 [0.016, 0.176] & 5 \\
\bottomrule
\end{tabular}
\end{table}

In validation, SAR and AC match 619 of 640 truths with no unmatched outputs; SC matches 615 and PC 619 with four unmatched outputs. SAR retains all 615 SC outputs and recovers four truths. It requests four of 157 weak reports using 80 cues from 40 anchors; equal-sized records give 1,811 versus 1,884 equivalents (3.87\% less). The paired miss reduction is 0.625 percentage points [0.156,1.250]; all 615 common truths retain identical position/velocity estimates. The conservative two-sided 95\% upper bound on validation mean FA2 is 2.841\%. All policies have zero alarms in 128 validation empty drops, with a one-sided 95\% upper bound of 2.313\%.

\balance
\section{Conclusion}
SAR preserves SC estimates while recovering rejected targets, matching AC detection counts in the evaluated populations. Against uploading all eligible reports, weak uploads fall 97.4\% and modeled bidirectional payload falls 4.5\% for equal cue/report records. Future work will optimize report/query encoding and scheduling.


\begin{thebibliography}{9}
\bibitem{wei} Z. Wei \emph{et al.}, ``5G PRS-based sensing: A sensing reference signal approach for joint sensing and communication system,'' \emph{IEEE Trans. Veh. Technol.}, vol. 72, no. 3, pp. 3250--3263, Mar. 2023.
\bibitem{shi} Q. Shi and L. Liu, ``Joint LOS identification and data association for 6G-enabled networked device-free sensing,'' \emph{IEEE Trans. Commun.}, vol. 72, no. 8, pp. 5117--5129, Aug. 2024.
\bibitem{tang} J. Tang \emph{et al.}, ``Cooperative ISAC-empowered low-altitude economy,'' \emph{IEEE Trans. Wireless Commun.}, vol. 24, no. 5, pp. 3837--3853, May 2025.
\bibitem{weighted} H. Zhang \emph{et al.}, ``A novel multilevel weighted fusion for ISAC-enabled cooperative AAV localization,'' \emph{IEEE Commun. Lett.}, vol. 29, no. 10, pp. 2273--2277, Oct. 2025.
\bibitem{perovic} N. S. Perovi\'{c}, M. F. Flanagan, and L.-N. Tran, ``Sensing rate optimization for multi-band cooperative ISAC systems,'' \emph{IEEE Commun. Lett.}, vol. 29, no. 9, pp. 2163--2167, Sep. 2025.
\bibitem{zhu} W. Zhu, S. Zhang, and L. Liu, ``Joint transmission and compression optimization for networked sensing with limited-capacity fronthaul links,'' \emph{IEEE Trans. Wireless Commun.}, vol. 24, no. 8, pp. 6643--6657, Aug. 2025.
\bibitem{jiang} R. Jiang and B. Chen, ``Fusion of censored decisions in wireless sensor networks,'' \emph{IEEE Trans. Wireless Commun.}, vol. 4, no. 6, pp. 2668--2673, Nov. 2005.
\bibitem{study} 3GPP, ``Study on integrated sensing and communication (ISAC) for NR,'' TR 38.765, Release 20, 2026.
\end{thebibliography}
\end{document}